\documentclass[conference]{IEEEtran}
\IEEEoverridecommandlockouts
\usepackage{cite}
\usepackage{amsmath,amssymb,amsfonts}
\usepackage{algorithmic}
\usepackage{booktabs}
\usepackage{cleveref}
\usepackage{graphicx}
\usepackage{textcomp}
\usepackage{xcolor}
\def\BibTeX{{\rm B\kern-.05em{\sc i\kern-.025em b}\kern-.08em
    T\kern-.1667em\lower.7ex\hbox{E}\kern-.125emX}}
\begin{document}

\title{A performance analysis of VM-based Trusted Execution Environments for Confidential Federated Learning\\
\thanks{This work has been partly supported by the Spoke "FutureHPC \& BigData” of the ICSC – Centro Nazionale di Ricerca in "High Performance Computing, Big Data and Quantum Computing", funded by European Union – NextGenerationEU, and by the Horizon2020 RIA EPI project (G.A. 826647).
We thank the "Intel\textsuperscript{\textregistered} SMG Customer Innovation Labs Swindon" laboratory for granting us access to their SGX-enabled machines.}
}


\author{\IEEEauthorblockN{1\textsuperscript{st} Bruno Casella}
\IEEEauthorblockA{\textit{University of Turin} \\
\textit{Alpha Research Group - Computer Science Department}\\
Turin, Italy \\
bruno.casella@unito.it \\
0000-0002-9513-6087 \\}
}

\maketitle

\begin{abstract}
Federated Learning (FL) is a distributed machine learning approach that has emerged as an effective way to address recent privacy concerns. However, FL introduces the need for additional security measures as FL alone is still subject to vulnerabilities such as model and data poisoning and inference attacks. Confidential Computing (CC) is a paradigm that, by leveraging hardware-based trusted execution environments (TEEs), protects the confidentiality and integrity of ML models and data, thus resulting in a powerful ally of FL applications. Typical TEEs offer an application-isolation level but suffer many drawbacks, such as limited available memory and debugging and coding difficulties. The new generation of TEEs offers a virtual machine (VM)-based isolation level, thus reducing the porting effort for existing applications. In this work, we compare the performance of VM-based and application-isolation level TEEs for confidential FL (CFL) applications. In particular, we evaluate the impact of TEEs and additional security mechanisms such as TLS (for securing the communication channel). The results, obtained across three datasets and two deep learning models, demonstrate that the new VM-based TEEs introduce a limited overhead (at most 1.5x), thus paving the way to leverage public and untrusted computing environments, such as HPC facilities or public cloud, without detriment to performance. 
\end{abstract}


\section{Introduction}\label{sec:intro}
Artificial Intelligence (AI) applications are reinventing daily tasks: AI chatbots, autonomous vehicles, and recommendation systems are becoming pervasive. 
AI is based on extracting patterns from data. However, real-world data is distributed due to its intrinsic nature. To train an AI model with good predictive capabilities, it is often required to aggregate data in a single data lake. However, due to recent privacy regulations, it is not always possible to share data. As a result, it is essential to guarantee the protection of sensitive data when building trustworthy AI models.\\
FL~\cite{mcmahan2017communication} is a distributed AI technique based on deep learning (DL) models such as neural networks (NN) that has emerged thanks to its privacy-preserving characteristics. The underlying idea of FL is to move DL models rather than data: in a typical federation, the distributed clients send their locally trained models to a central server that creates a global model by aggregating them, and the process is repeated. The first proposed FL algorithm is FedAvg~\cite{mcmahan2017communication}, in which the aggregation function is the coordinate-wise averaging of the model parameters.
While FL offers a high degree of trustworthiness that can be additionally boosted with techniques such as differential privacy~\cite{dwork2014differential}, it raises new security and learning challenges. For instance, the clients must trust the central aggregator, which can attempt to reconstruct original data from updates~\cite{evans2018pragmatic}, and vice-versa, as the participants can tamper with the training process, for example, by sending wrong updates.\\
CC gives the opportunity to exploit untrusted environments, such as the public cloud, to execute safe workloads on sensitive data.
CC and FL are fated to meet as CC is not a competing technology but a powerful helper of FL applications. 
CC leverages TEEs to move part of the process into a secure ``enclave'', whose code can be attested and verified. TEEs come with some features that make them suitable for establishing confidence that code has been run securely and faithfully. They provide confidentiality (the execution state of the code is kept hidden), integrity (the code's execution can not be affected), and attestation (it can demonstrate to a remote party what binary code is running and what its initial state was). Most companies in the manufacturing of processors have proposed their TEE. For example, Intel developed Intel Software Guard Extensions (Intel SGX)~\cite{anati2013innovative}, while ARM first proposed TrustZone~\cite{alves2004trustzone}. These technologies are a set of CPU instructions providing an application-isolation level. However, they come at the cost of heavy overhead and debugging difficulties, and they do not offer GPU compatibility.
The new generation of TEEs, such as the Intel TDX, ARM CCA, and AMD SEV-SNP, aims to overcome the main limitations of the old TEE architectures.\\
In this work, we benchmark the usage of the recently released Intel TDX machines to run robust and secure CFL applications. 
The main contributions of our work are: 1) we measure and compare the overhead introduced by Intel TDX with respect to the performance of Intel SGX, 2) we perform extensive CFL experiments on three image classification datasets and two different NN, 3) we show that VM-based TEEs have a little impact on the runtime overhead, thus paving the way to CC to become the default mode to deploy AI models, especially nowadays, that with the advent of large language models (LLMs), there is a need to rely on untrusted HPC and cloud facilities.

\section{Related Work and Methodology}\label{sec:related}
The recent advent of LLMs based on transformers~\cite{vaswani2017attention}, such as GPT~\cite{openai2023gpt} or Llama~\cite{touvron2023llama}, is increasing the adoption of third-party computing and storage resources, such as public cloud services or HPC facilities. Besides centralized AI deployment, FL applications are also starting to benefit from a vast amount of hardware availability. A recent work~\cite{colonnelli2023mlastro} investigated the potential of hybrid workflow models to train an FL pipeline across heterogeneous infrastructures with limited connectivity between worker nodes, showing that the overhead introduced by using a general-purpose workflow system does not significantly affect the execution time. Another work introduced XFFL, a cross-facility FL framework based on hybrid workflows, that trains a full Llamav2 instance on two facilities of the EuroHPC JU, showing how the increased computing power completely compensates for the additional overhead introduced by two data centers.
This shift towards utilizing cloud and HPC resources introduces unique security and reliability concerns, especially when combined with FL's inherent challenges. 
FL systems are susceptible to issues such as membership inference attacks~\cite{shokri2017mia}, in which the attacker aims to identify if a sample is in the training set, property inference attacks~\cite{ganju2018pia}, in which the goal is to infer a global private property of the data, and inversion/extraction attacks that aim to extract representative or complete examples from the training data~\cite{fredrikson2015inversion}.
As a result, securing these deployments has become essential, particularly to safeguard sensitive data and model integrity across infrastructures.\\ 
CC and TEEs address some of these concerns by providing isolated and encrypted environments for computation, ensuring that sensitive data remains protected throughout the learning process. These technologies enhance the privacy and reliability of FL systems by securely managing data and model updates across third-party resources, reinforcing trust in cloud-based and distributed FL deployments.
For these reasons, CFL is likely to become the default mode for deploying FL workloads~\cite{guo2024cfl}.\\
However, the first generation of TEEs comes with several drawbacks, such as 1) limited protected memory, 2) restriction to CPU resources (incompatibility with GPUs), 3) significant challenges in terms of code porting and adaptation, and 4) runtime overhead. 
PPFL~\cite{ppfl2021mo}, a recent work leveraging Arm TrustZone for CFL applications, addressed the memory constraint by adopting a layer-wise training strategy to train layers one at a time. However, it introduces a 3x runtime delay and affects the learning performance.
Other works adopt trusted containers, which can be library OSes or libc wrappers, to mitigate these issues. For instance, SecFL~\cite{secfl2021lequoc} performs training inside TEE enclaves by leveraging the SCONE~\cite{scone2016arnautov} library OS. 
Another recent work~\cite{casella2024analysis} proposes a performance analysis of CFL for image classification tasks by evaluating the overhead of different security mechanisms to protect the data and model during computation (TEE), communication (TLS~\cite{tls}), and storage (disk encryption). The authors leveraged the Gramine~\cite{graphene2014tsai} libOS due to its compatibility with the OpenFL~\cite{openfl2022foley} framework for running a real distributed CFL pipeline. Both the server and the clients instances are executed in dedicated SGX-enabled machines using Gramine. For easy of deployment they encapsulated a graminized version of OpenFL inside Docker images, thus providing a high level of security and portability. The authors found empirically that an accurate phase of system tuning is required to avoid undesired delays in wall-clock times: the best runtime performance are achieved when setting the number of threads equal to the number of cores per socket of the running machine. Results show that library OSes represent a practical way to inject security measures into FL without incurring into prohibitive overheads (up to 2x).\\
In this work, we benchmark the usage of the new generation TEEs for executing CFL applications. In order to compare with the performance of application-level TEEs~\cite{casella2024analysis}, we relied on the Trust Domain eXtensions (TDX), the new family of VM-based TEEs developed by Intel.
Intel TDX is a new architectural extension present in some Intel 4th generation Xeon processors, and generally available in 5th generation Emerald Rapids, and it offers VM-level isolation. It is designed to facilitate the deployment of trust domains (TD), hardware-isolated VMs developed to protect sensitive data and code from unauthorized accesses. A CPU-measured module enables Intel TDX. This software module operates in the new CPU Secure Arbitration Model (SEAM) as a peer VM manager (VMM). This allows the TD in the entry/exit operations using the existing virtualization infrastructure. The module is hosted in a reserved memory space identified by the SEAM Range Register. TDX uses hardware extensions for managing and encrypting memory and protects both the confidentiality and integrity of the TD CPU state from non-SEAM mode.
Additionally, Intel TDX incorporates several architectural elements such as a secure Extended Page Table (EPT), and the Intel Total Memory Encryption – Multi-Key (Intel TME-MK) to ensure robustness and security.\\
We also adopted the Transport Layer Security (TLS) to encrypt the communication between clients and server over the network. TLS adds overhead due to its processes of handshake, derivation of the encryption keys and to encrypt/decrypt the data when sending or receiving over the secured channel.

\section{Experiments and Discussion}\label{sec:experiments}
The goal of our experiments is to study the runtime performance of a CFL application leveraging Intel TDX and TLS to secure computation and communication and to compare it against a baseline~\cite{casella2024analysis} built on an application-isolation level TEEs such as Intel SGX. We use the following experimental protocol.\\
We use OpenFL for training our image classification models. In OpenFL, the server and the clients are called Aggregator and Collaborators. For each federation round, each collaborator executes two validation steps (one for the received global model, and one for the locally trained model) and a training stage (in which the global model is trained on local and private data).\\
All experiments used a distributed environment encompassing one Aggregator and three Collaborators, each deployed on a dedicated server.
All TDX-enabled servers are dual sockets with two 5th generation Intel® Xeon Platinum 8592 CPUs (64 cores @ 2.00GHz). The servers communicate via 100Gb/s Ethernet links. As software, we used OpenFL v1.5.
The competitors run on dual sockets with two 3$^{\text{rd}}$ generation Intel\textsuperscript{\textregistered} Xeon Scalable Platinum 8380 CPUs (40 cores @ 2.30GHz) with SGX support, with a 100Gb/s Ethernet connection, OpenFL v1.5 and Gramine v1.4.\\
To compare with the baseline, we reproduced their learning scenario. We trained two models, a ResNet-18~\cite{resnet2016he} and a MobileNetV3-Small~\cite{mobilenet2019howard}, for 100 rounds using the cross-entropy loss, the Adam Optimizer, and a learning rate of $1e^{-4}$, on three image classification benchmark datasets, i.e., MNIST~\cite{lecun1998gradient}, CIFAR10 and CIFAR100~\cite{krizhevsky2009learning}.

The statistics of datasets and models for the CIFAR10 (CIFAR100 slightly increases the number of parameters and model size by 0.41\% due to bigger size of the output layer to handle more classes, while colours MNIST slightly decreases them by 0.06\% due to smaller inputs/fewer channels) are summarized in~\Cref{tab:details}.

\begin{table}[t]
\centering
\begin{tabular}{lccccc} 
\hline \\[-1.0em]
\textbf{} & \textbf{MNIST} & \textbf{CIFAR10} & \textbf{CIFAR100} \\ 
\hline \\[-1.0em]
Dimension & 28x28 & 32x32 & 32x32\\
Representation & Binary & RGB & RGB\\
Train samples & 60.000 & 50.000 & 50.000\\
Test samples & 10.000 & 10.000 & 10.000\\
Size (MB) & 64 & 170 & 169\\
\hline \\[-1.0em]
\end{tabular}
\\[1.0em]
\centering
\begin{tabular}{lccc} 
\hline \\[-1.0em]
\textbf{} & \textbf{ResNet-18} & \textbf{MobileNetV3-Small} \\ 
\hline \\[-1.0em]
Parameters & 11.181.642 & 1.528.106\\
Size (MB) & 42.69 & 5.88\\
Layers & 68 & 210\\
\hline \\[-1.0em]
\end{tabular}
\caption{Statistics of datasets and models.}
\label{tab:details}
\end{table}

We train the two models on the three datasets, starting from the baseline and incrementally adding security mechanisms, i.e. the TEE and secure communications (TLS). Each one adds some extra execution steps. ``Gramine-direct'' redirects the system calls through Gramine. SGX further runs the application in the security enclave.

\begin{table}[!ht]
\centering
\resizebox{0.49\textwidth}{!}{%
\begin{tabular}{lcccccccccc}
\hline \\[-1.0em]
\textbf{} & \multicolumn{3}{c}{\textbf{ResNet-18}} & \multicolumn{3}{c}{\textbf{MobileNetV3-Small}} \\
\cmidrule(r){2-4} \cmidrule(l){5-7}
 & \textbf{MNIST} & \textbf{CIFAR10} & \textbf{CIFAR100} & \textbf{MNIST} & \textbf{CIFAR10} & \textbf{CIFAR100} \\
\hline \\[-1.0em]
Baseline & 01:00 & 01:10 & 01:00 & 00:53 & 00:57 & 00:58 \\
Baseline + TLS & 01:04 & 01:12 & 01:08 & 00:54 & 00:59 & 01:02 \\
TDX  & 01:05 & 01:18 & 01:11 & 01:16 & 01:07 & 01:07 \\
TDX + TLS  & 01:08 & 01:26 & 01:15 & 01:18 & 01:11 & 01:12\\
\hline \\[-1.0em]
\end{tabular}}
\\[1.0em]
\resizebox{0.49\textwidth}{!}{%
\centering
\begin{tabular}{lcccccccccc}
\hline \\[-1.0em]
\textbf{} & \multicolumn{3}{c}{\textbf{ResNet-18}} & \multicolumn{3}{c}{\textbf{MobileNetV3-Small}} \\
\cmidrule(r){2-4} \cmidrule(l){5-7}
 & \textbf{MNIST} & \textbf{CIFAR10} & \textbf{CIFAR100} & \textbf{MNIST} & \textbf{CIFAR10} & \textbf{CIFAR100} \\
\hline \\[-1.0em]
Baseline & 01:55& 02:01& 01:54& 01:40& 01:38& 01:29\\
\textit{Gramine-direct} & 03:10& 03:33&  03:32& 02:12& 01:54& 02:03\\
SGX  & 03:17& 03:36& 03:49& 02:43& 02:24& 02:36\\
SGX + TLS  & 03:42& 03:53&  03:59& 02:53& 02:37& 02:55\\
\hline \\[-1.0em]
\end{tabular}}

\caption{Wall-clock execution times [HH:MM] to train models on 100 federated training rounds of 1 epoch.}
\label{tab:wall-clock}
\end{table}

Table~\ref{tab:wall-clock} summarizes the total wall clock times of each experiment.
Results expressed in the HH:MM time format reveal an easily discernible pattern. As expected, each addition of a security measure leads to an increase in execution time. 
However, while for SGX it is possible to state that a fully encrypted FL pipeline doubles the execution time, TDX adds only a slight overhead to the wall-clock time.

Since our approach is not affected by loss in learning performance, as we perform a standard training routine as in~\cite{casella2024analysis}, and due to space constraints, we omit plots and discussion about model's accuracies.

To better understand the overhead added by each single mechanism, we reproduced the performance model presented in~\cite{casella2024analysis}. The execution time $T$ can be seen as a linear combination with one term to capture the relationship with each discussed security mechanism:
\begin{equation}
    T = T_{baseline} + O_{TDX} + O_{TLS}\notag
\end{equation}

where $T_{baseline}$ is the runtime without any security mechanism and $O_{TDX}, O_{TLS}$ represent the overheads introduced by TDX and TLS, respectively. 

We express each overhead as a cost coefficient $C$ for the $T_{baseline}$, i.e. $O = C \cdot T_{baseline}$. Considering that both TDX and TLS only add extra time, we fit a non-negative linear regression model. Table~\ref{tab:coefficients} shows the fitted values for both NNs.

\begin{figure*}
  \begin{minipage}{0.49\linewidth} 
  \centering
    \begin{tabular}{c}
        \\
        \text{$O_{TDX}$} = \text{$C_{TDX} \cdot T_{baseline}$} \notag\\
        \text{$O_{TLS}$} = \text{$C_{TLS} \cdot T_{baseline}$} \notag
        \end{tabular}
    \caption{Overhead terms of the wall-clock times.}
    \label{fig:overhead_terms}
  \end{minipage}
  \hfill 
  \begin{minipage}{0.49\linewidth} 
    \centering
    \footnotesize
    \begin{tabular}{lcc}
        \toprule
                    & ResNet-18 & MobileNetV3-Small \\
        \midrule
        $C_{TDX}$         & 0.126     & 0.245             \\
        $C_{TLS}$         & 0.079     & 0.065             \\
        \bottomrule
    \end{tabular}
    \caption{Coefficients of the linear regression models.}
    \label{tab:coefficients}
  \end{minipage}
\end{figure*}

The overhead introduced by TDX is minimal with respect to the delay induced by SGX (managed by Gramine). This happens because TDX deletes the constant switches between the enclave and the unprotected world, as in SGX, at the cost of slightly decreasing the security measures. Indeed, while SGX provides the smallest attack surface (only the enclave can access the confidential data), TDX increases the trust boundaries to guest OS, all the applications, and VM admins. However, TDX can be more amenable to enterprise-wide deployment scenarios thanks to its lower porting effort for existing applications.
Surprisingly, the overhead coefficient of MobileNetV3-Small is double the ResNet-18 coefficient. The author hypothesizes that this can be due to some optimization reasons. For instance, MobileNetV3-Small operates with particular operations, the depthwise separable convolutions, that are specifically devised for reducing the number of parameters of the network, but they may not be optimized in modern hardware such as TDX. Moreover, MobileNetV3-Small has a larger number of layers, about 3.1 times more than ResNet-18, which, due to the internals of OpenFL, increases the number of I/O and network operations.

\begin{table}[!ht]
\centering
\resizebox{0.49\textwidth}{!}{%
\centering
\begin{tabular}{lcccccccccc}
\hline \\[-1.0em]
\textbf{} & \multicolumn{3}{c}{\textbf{ResNet-18}} & \multicolumn{3}{c}{\textbf{MobileNetV3-Small}} \\
\cmidrule(r){2-4} \cmidrule(l){5-7}
 & \textbf{MNIST} & \textbf{CIFAR10} & \textbf{CIFAR100} & \textbf{MNIST} & \textbf{CIFAR10} & \textbf{CIFAR100} \\
\hline \\[-1.0em]
Baseline & 2.19 $\pm$ 0.40 & 1.13 $\pm$ 0.34 & 1.24 $\pm$ 0.43 & 1.20 $\pm$ 0.40 & 2.31 $\pm$ 0.46 & 2.33 $\pm$ 0.47 \\
Baseline + TLS & 3.46 $\pm$ 0.50 & 1.88 $\pm$ 0.32 & 1.89 $\pm$ 0.32 & 1.85 $\pm$ 0.36 & 3.66 $\pm$ 0.48 & 3.59 $\pm$ 0.49 \\
TDX & 1.76 $\pm$ 0.43 & 1.83 $\pm$ 0.38 & 1.82 $\pm$ 0.39 & 3.27 $\pm$ 0.63 & 3.30 $\pm$ 0.64 & 3.19 $\pm$ 0.39 \\
TDX + TLS & 2.60 $\pm$ 0.49 & 2.69 $\pm$ 0.46 & 2.61 $\pm$ 0.49 & 4.81 $\pm$ 0.40 & 4.87 $\pm$ 0.34 & 4.80 $\pm$ 0.41 \\
\hline \\[-1.0em]
\end{tabular}}
\caption{Communication times in seconds (mean $\pm$ standard deviation of three clients and 100 federated training rounds).\\}
\label{tab:communication_times-tdx}
\end{table}

Finally, from a communication point of view, both TDX and TLS add overhead, except ResNet-18 on MNIST, which may be due to some internal factors of that particular run. As expected, the greater source of delay is given by TLS, which adds the handshake process and key derivation processes for securing the network. The absolute times align with the communication times of the SGX experiments~\cite{casella2024analysis}: MobileNetV3-Small requires approximately double the amount of communication time required by ResNet-18. This happens because OpenFL requests the global model layer by layer, and MobileNet-V3 has roughly 3x the layers of ResNet-18 (even if they are smaller).

\section{Conclusion}\label{sec:conclusion}
In this work, we extended the performance model for CFL proposed in a recent work~\cite{casella2024analysis} adopting Intel SGX with the new VM-based TEEs Intel TDX. We conducted extensive experiments and analyzed the overhead introduced by TDX and TLS. Results spanning over three datasets and two models show that CC can become the standard model to deploy AI models in untrusted facilities without incurring undesired overhead.

\vspace{12pt}

\end{document}